  \documentstyle[amssymb,aps,multicol,psfig,eqsecnum]{revtex}  
  \input epsf

\begin{document}
\draft
\title{Early Stages of Homopolymer Collapse}
\author{A. Halperin}
\address{UMR 5819 (CEA-CNRS-UJF), DRFMC, 
 \\
CEA-Grenoble, 17 rue des Martyrs, 
 \\
38054 Grenoble Cedex 9, France}
\author{Paul M. Goldbart}
\address{Department of Physics, 
 \\
University of Illinois at Urbana-Champaign, 
 \\
1110 West Green Street, 
 \\
Urbana, Illinois 61801-3080, U.S.A.}
  \date{May 20, 1999}
\maketitle
\vskip 2truein
\begin{abstract}
\noindent
{\it Abstract\/}: 
Interest in the protein folding problem has motivated a wide range of 
theoretical and experimental studies of the kinetics of the collapse 
of flexible homopolymers.  In this Paper a phenomenological model is 
proposed for the kinetics of the early stages of homopolymer collapse
following a quench from temperatures 
above to below the $\theta$ temperature.  In the first stage, nascent 
droplets of the dense phase are formed, with little effect on the 
configurations of the bridges that join them.  The droplets then grow 
by accreting monomers from the bridges, thus causing the bridges to 
stretch.  During these two stages the overall dimensions of the chain 
decrease only weakly.  Further growth of the droplets is accomplished 
by the shortening of the bridges, which causes the shrinking of the 
overall dimensions of the chain. The characteristic times of the three
stages respectively scale as $N^{0}$, $N^{1/5}$ and $N^{6/5}$, where 
$N$ is the degree of polymerization of the chain.
\end{abstract}
\vfil\eject
\begin{multicols}{2}
\section{Introduction}

The purpose of this Paper is to discuss the evolution of the structure of a
single polymer molecule after the environment has been changed from a good
solvent to a poor solvent. Our approach is via a phenomenological model for
the kinetics of the early stages of collapse. We have in mind a long
flexible polymer chain immersed in a simple solvent, and a change in the
temperature $T$ as the means of altering the solvent quality.\ The initial
state is one of equilibrium in a good solvent, in which the chain is swollen
and has radius of gyration $R=R_{F}\thicksim N^{3/5}$, where $N$ is the
degree of polymerization. The final state is one of equilibrium in a poor
solvent, in which the chain is collapsed into a dense, spherical globule, 
i.e., $R=R_{c}\thicksim N^{1/3}$.\ The variation of the equilibrium
state with the solvent quality is, with some important caveats, well
understood~\cite{PGG,GK,JdC,WBF}. 
In marked contrast, there
is no comparable consensus concerning the kinetics of collapse following an
abrupt $T$ quench, i.e., a rapid change in temperature from above to
below the $\theta $ temperature. The experimental picture is not clear
because of difficulties due to the competition between the collapse of the
individual chains and the aggregation of different ones. As a result, there
have been only few experimental studies of the kinetics of collapse of
isolated chains~\cite{Chu,Chi,Nakata}. 
Theoretical studies have
utilized a variety of approaches in focusing on different aspects of the
problem. Langevin models~\cite{Dawson,Allegra,Garel},
phenomenological models~\cite{PGG1,Beguin,Klushin} 
and computer
simulations~\cite{KS,Byme,Mattice,Dawson1,Wittkop,Ostrovsky,Milchev} 
have been used to consider the kinetics of
collapse in the absence of topological constraints. Other studies have
focused on the role of internal entanglements~\cite{GNS,RGT} 
or the
competition with chain-chain aggregation~\cite{GKu,Raos}. 
In spite of
the extensive activity is this area, a complete picture of the kinetics of
collapse has yet to emerge. The search for such picture is motivated
primarily by the intense current interest in the protein folding 
question~\cite{folding,Pande,EIS} 
because early stages of protein folding
are thought to proceed as the collapse of flexible homopolymers.

\begin{figure}[hbt]
 \epsfxsize=3.5in
 \centerline{\epsfbox{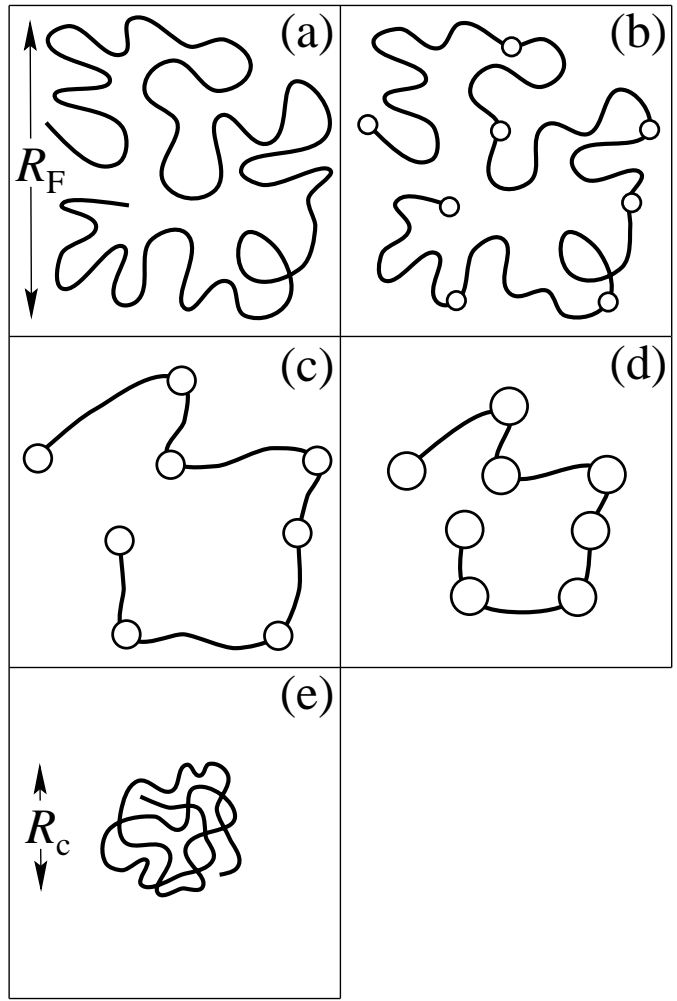}}
\vskip0.5truecm
\caption{Proposed sequence of states during the collapse of 
\\ 
a flexible homopolymer: 
\\ 
(a)~self-avoiding configuration under good solvent conditions; 
\\ 
(b)~pearling; 
\\ 
(c)~bridge stretching; 
\\
(d)~collapse of the pearl necklace; 
\\
(e)~equilibrium collapsed configuration in a poor solvent.}
\label{FIG:figone}
\end{figure}

The collapse of a flexible chain following an abrupt change of solvent
quality displays certain features of a first-order phase transition. These
features occur for chains of finite $N$ for which the surface free energy of
the globule plays an important role. Two such features are important to the
present discussion. First, the onset of collapse involves, as we shall
discuss, the formation of droplets of the dense ``phase''. Second, the
collapse proceeds via intermediate states corresponding to the ``tadpole''
configuration of a stretched, collapsed globule~\cite{HZ}, which involve the
coexistence of dense globules and strongly stretched segments. With this
picture in mind, it is possible to distinguish between four stages in the
kinetics of collapse (Fig.~\ref{FIG:figone}). 
Each stage is characterized by a
length-scale and a time-scale. The characteristic time scale increases with
the relevant length scale. The formation of $p$ ($\ll N$) droplets of the dense
phase, which we refer to as ``pearls'', is the fastest process. This
``pearling'' stage involves local rearrangement of the chain configurations
within the randomly placed droplets during which the dimensions and
configurations of the chain as a whole are only weakly modified. The 
\noindent
nascent
droplets then grow by accreting monomers from the bridges that connect them.
The positions of the growing droplets are roughly stationary and their
growth is accommodated by the gradual stretching of the bridges. 
This ``bridge-stretching'' stage ends 
when the tension in the bridges attains the
equilibrium value $f_{co}$ corresponding to the ``tadpole'' configuration of
a stretched globule. During this process the dimensions of the chain as a
whole remain essentially constant {\em i.e,} $R\sim N^{3/5},$ but the
configurations of the bridges are qualitatively modified. The number of
droplets $p$ is set by the requirement that $R\sim N^{3/5}$ at the end of
this stage, thus leading to $p\sim N^{4/5}$. Beyond this point, the growth
of the ``pearls'' is accompanied by shortening of the bridges and the
overall shrinking of the chain. The third stage, the collapse of the ``pearl
necklace'', involves the chain as a whole. The corresponding length scale $%
R_{F}(\sim N^{3/5})$ is larger and the characteristic time is thus longer.
It is assumed that $p$ remains roughly constant during these three stages.
Eventually, the droplets come into contact, and coalesce into a single
globule. Our discussion is limited to the early stages of the collapse. The
last regime will not be considered in the present paper. Within this
picture, the characteristic times for the first three stages, pearling,
bridge-stretching and the collapse of the pearl necklace are, respectively, $%
\tau _{p}\sim N^{o}$, $\tau _{bs}\sim N^{1/5}$ and $\tau _{pn}\sim N^{6/5}$.
The model we propose differs from the existing phenomenological descriptions
in two respects. First, the two initial stages were not considered in these
models. Second, the treatment of the third stage, the collapse of the pearl
necklace, is different. A more detailed comparison with the phenomenological
models, as well as the Langevin-type theories and to simulation studies,
will be presented in the final Section.

The Paper is organized as follows. To obtain the temporal evolution of the
various stages we balance the driving force with the entropy production. A
discussion of this approach is presented in Section II. In Section III we
recall some key results concerning collapse blobs and the thermodynamics of
collapse. In addition, we discuss the initial role of the collapse blobs as
spinodal ``modes'' and the equilibrium deformation behavior of a collapsed
chain. The collapse of a chain from a good solvent to a $\theta $ solvent is
discussed in Section IV. The results are then used to analyze the first
stage of the collapse, i.e., the formation of nascent droplets or
``pearling''. In Section V we discuss the collapse of a chain with
constrained ends. This discussion is then utilized to describe the second
stage of collapse, in which the stationary droplets, pearls, grow by
accreting monomers from the bridges thus causing the bridges to stretch. The
shrinking of the resulting string of pearls is analyzed in Section VI.

\section{Some General Considerations; Entropy Production Method}

The design of the phenomenological model presented in this paper, involves
two steps. In the first step we propose a sequence of stages followed by the
collapsing chain (The stages involved were outlined in the introduction.)
The second step is to calculate the characteristic time associated with the
various stages in the proposed route. A simple method for calculating the
characteristic times was proposed by de Gennes~\cite{PGG1}. It focuses on a
volume element of solvent containing a single chain and coupled to a thermal
reservoir that maintains a temperature $T$. The systems as a whole, is
closed. During collapse, the free energy of the chain $F_{chain}$ decreases.
The decrease in $F_{chain}$ gives rise to an increase in $S_{total}$ the
total entropy of the system as a whole. Since the collapse is the only
irreversible process involved the changes in $F_{chain}$ and $S_{total}$ are
related by 
\begin{equation}
\Delta F_{chain}=-T\Delta S_{total}.  \label{ii1}
\end{equation}
The dynamics of the process are described by 
\begin{equation}
\frac{dF_{chain}}{dt}=-T\frac{dS_{total}}{dt}.  \label{ii2}
\end{equation}
To implement this argument, it is necessary to specify: (i) $F_{chain}$ as a
function of the chain radius $R$. One assumes that $dF_{chain}/dt=$ $%
(dF_{chain}/dR)V,$ where $V=dR/dt$. The form of $F_{chain}$ for the various
stages will be discussed at the appropriate sections. One must also specify
(ii) the operative dissipation mechanisms that determine $dS_{total}/dt$ in
terms of $R$ and $V$. This entropy production approach is independent of the
model, in that its applicability is not limited to a specific choice of $%
F_{chain}$ or of the dissipative modes. Nor does it provide guidance for
choosing this crucial input.

Two dissipation mechanisms are considered for the various stages of
collapse. First is the viscous dissipation associated with the motion of the
solvent. Second is the frictional dissipation due to the Stokes drag on the
chain. The magnitude of the viscous force~\cite{Hydro} acting on a unit
volume is on the order of $\eta \nabla ^{2}V\,$where $\eta $ is the shear
viscosity. In the following we estimate it by $\eta V/R^{2}.$\ The
associated dissipation per unit time is thus roughly $\eta V^{2}/R^{2}$ and
the total hydrodynamic dissipation within a volume $R^{3}$ is $TdS_{h}$ as
given by 
\begin{equation}
T\frac{dS_{h}}{dt}\approx \eta \frac{V^{2}}{R^{2}}R^{3}\approx \eta V^{2}R.
\label{ii3}
\end{equation}
The Stokes drag force acting on a sphere of radius $r$ moving with a
constant velocity $V$ is $6\pi \eta Vr,$\ and the associated ``frictional''
dissipation per particle is 
\begin{equation}
T\frac{dS}{dt}\approx \eta V^{2}r.  \label{ii4}
\end{equation}
To obtain the total frictional dissipation it is necessary to sum the
contributions due to the hydrodynamically impenetrable objects involved in
the stage considered. The number of such objects, their radius $r$ and the
relationship between $r$ and $R$ differ in the various stages of collapse.
The de Gennes argument is equivalent to equating the frictional force to the
contraction force. It allows us however, to consider on equal footing the
dissipation due to Stokes drag forces and to the hydrodynamic flows.
Finally, note that the dissipation increases with the length scale. This
observation motivates the proposed sequence of stages which involve
processes taking place at progressively larger length scales.

The familiar methods of irreversible thermodynamics~\cite{Callen} focus on
linear responses, the Onsager relations, and their refinements. In marked
difference, the de \ Gennes approach is not limited to a linear regime nor
does it relay on the analysis of the regression of fluctuations. The method
applies to the complete dynamic range, provided that the driving forces and
the dissipation mechanisms are correctly identified. The dissipative modes
invoked in our discussion hold under the assumption that the hydrodynamic
description is applicable to the collapse of a polymer. It is expedient to
use this approach, but one should note that the validity of continuum
theories at this length scale is a delicate issue.

\section{Collapse Blobs, Spinodal Decomposition and the Extension Elasticity
of Globules}

Whilst the configurations of polymers in good and poor solvents are well
understood, the nature of the transition between the two regimes is not
fully resolved. Different models suggest that the transition may be
continuous or discontinuous, depending on $N$ and the stiffness of the chain.
The issue of the order of the transition is, however, irrelevant when
considering a collapse following a quench to the poor solvent regime. In
this regime, the equilibrium configuration of the chain is determined
primarily by the Flory interaction free energy $F_{int}$. The elasticity of
the chain, i.e., its configurational entropy, plays a relatively minor
role, limited to determining the structure of 
the globule-solvent interface~\cite{GK}. 
In the case of a single chain, for which the translational
entropy of the polymer is irrelevant, the $N\rightarrow \infty $ limit of $%
F_{int}$ determines the equilibrium behavior: 
\begin{equation}
\frac{F_{int}}{kT}=(1-\phi )\ln (1-\phi )+\chi \phi (1-\phi ).  \label{iii1}
\end{equation}
Here, $\phi $ is the monomer volume-fraction and $\chi $, the Flory
parameter, is a measure of the strength of the interactions giving rise to
the mixing enthalpy. For a given polymer-solvent system, $\chi $ depends
only on the temperature $T.$ In the good solvent regime, for which $\chi
<1/2,$ $F_{int}$ is a convex function of $\phi $. A critical point occurs at 
$(\chi _{c},$ $\phi _{c})$ specified by $\chi _{c}=1/2$ and $\phi _{c}=0.$
When $\chi >1/2$ the solvent is poor and the plot is concave in the range $%
0<\phi <\phi _{+}.$ A common tangent construction between the origin and $%
\phi _{+}$ indicates the coexistence of a dense phase, of concentration $%
\phi _{+}$ and the neat solvent. The $\phi _{+}$ phase corresponds to the
collapsed globule. This amounts to setting the osmotic pressure of the
polymer $\pi $ within the dense phase to zero, as is expected because there
are no free polymers in the solvent. The vanishing of $\pi $ leads to $\ln
(1-\phi )=\phi +\chi \phi ^{2}$.\ In the vicinity of the $\theta $
temperature, the $\ln (1-\phi )$ term may be approximated by the first three
terms of its Taylor expansion. The $\pi =0$ requirement then leads to 
\begin{equation}
\phi _{+}\approx -v/2w,  \label{iii2}
\end{equation}
where $v=a^{3}(1-2\chi )<0$ is the second virial coefficient, $a$ is the
size of the monomer, and $w=1/6>0$ is the third virial coefficient. Thus, $%
\phi _{+}$ is determined by the balance of attractive binary monomer-monomer
interactions and higher-order repulsive interactions. The chain elasticity
does not play a role. In turn, $\phi _{+}$ sets the radius of the collapsed
globule via $Na^{3}/R^{3}\approx \phi _{+},$ leading to $R/a\approx (N/\phi
_{+})^{1/3}.$ The occurrence of an unstable region in $F_{int}(\phi )$ is a
crucial feature of this discussion. Due to this feature, the collapse of a
quenched chain exhibits certain characteristics of first-order phase
transition.

The structure of the interface, and the corresponding surface tension, can
not be obtained from the discussion presented above. A ``blobological''
argument can provide the missing information~\cite{WBF}. The configurations
of a collapsed chain are characterized by a correlation length, $\xi _{c}.\,$%
\ On length scales smaller than $\xi _{c}$ the monomer-monomer attraction is
too weak to perturb the configurations of the chain, and it behaves as a
random walk. This allows one to define collapse blobs, which comprise $g_{c}
$ monomers such that $g_{c}^{1/2}a\thickapprox \xi _{c}.$ A fully collapsed
chain can be envisioned as a spherical globule composed of close-packed $\xi
_{c}$-blobs. The correlation length $\xi _{c}$ is determined by equating the
interaction free-energy due to monomer-monomer attraction in such a chain
segment to $kT.$ The interaction free-energy is $vg_{c}^{2}/\xi _{c}^{3},$
where $v$ is the second virial coefficient. In the vicinity of the $\theta $
temperature, $v$ may be expressed as $v\approx a^{3}(\Delta T/\theta ),$
where $\Delta T\equiv (T-\theta )/\theta ,$ thus leading to 
\begin{equation}
g_{c}\approx \left( \frac{\theta }{\Delta T}\right) ^{2}, 
\qquad 
\xi_{c}/a\approx \left( \frac{\theta }{\Delta T}\right) .  
\label{iii3}
\end{equation}
The density $ga^{3}/\xi _{c}^{3}$ within a single blob is comparable to that
of the dense phase $\phi _{+}$ resulting from the\ phase separation of free
polymers in a poor solvent. The $\xi _{c}$-blobs attract each other, and the
unperturbed chain forms a collapsed globule of radius $R\approx
(N/g_{c})^{1/3}\xi _{c}.$ A collapse blob in contact with the solvent is
assigned an energy of $kT.$ Thus, the boundary of the globule is associated
with a surface free energy 
\begin{equation}
\gamma \approx kT/\xi _{c}^{2}.  \label{iii4}
\end{equation}
Note that the blob picture, as described above, is valid only when the chain
comprises many collapse blobs, i.e., when the depth of the quench is
such that $\theta \gg \mid \Delta T\mid \gg \theta /N^{1/2}$ and $1\ll
g_{c}\ll N$. We should also add that in the following we treat the $\xi _{c}$%
-blobs as hydrodynamically impenetrable spheres.

When $N$ is sufficiently large, the collapsed globule may be viewed as a
droplet of the dense phase with density $\phi _{+}$ in coexistence with the
neat solvent. Within our model, the first stage of collapse involves $\phi
_{+}$ droplets, pearls, that are joined by bridges which retain the
configurations found in good solvent conditions. This situation is
reminiscent of the condensation of a fluid droplets from a supersaturated
vapor~\cite{Ma}. In this last situation it is found that droplets whose
radius is smaller than a critical radius $r_{c}$ tend to evaporate whereas
larger drops, i.e., those for which $r>r_{c},$ grow indefinitely. This
similarity wrongly suggests that the formation of nascent pearls is
analogous to the nucleation of critical droplets. This analogy is
discredited by the following rough argument. The thermodynamic potential of
a dense droplet coexisting with ``free'' monomers is given by 
\begin{equation}
\Lambda =F(N_{d})-N_{d}\mu ,  \label{iii5}
\end{equation}
where $N_{d}$ is the number of monomers in the droplet, and $\mu $ is the
chemical potential of the free monomers. For a sufficiently large droplet $%
F=N_{d}\mu _{d}+\gamma r^{2},$ where $\mu _{d\text{ }}$ is the chemical
potential of a monomer within the droplet. Choosing the reference state to
be a blob within the droplet and assigning $kT$ to an ``uncondensed'' blob,
leads to 
\begin{equation}
\frac{\Lambda }{kT}\approx -\frac{N_{d}}{g_{c}}+\left( \frac{N_{d}}{g_{c}}%
\right) ^{2/3},  \label{iii6}
\end{equation}
which has a maximum at $N_{d}\approx g_{c}.$ As Eq. (\ref{iii6}) is only
valid for $N_{d}$ $\gg g_{c},$ this establishes only that there is no
critical droplet of larger size. This argument does however suggest that
chain segments incorporating $g_{c}$ monomers play a special role in the
onset of collapse.\ This view is consistent with the definition of a $\xi
_{c}$-blob: The monomer-monomer attraction is too weak to perturb the chain
configurations on smaller length scales, but is sufficient to induce
collapse on length scales larger than $\xi _{c}.$ It is also helpful to note
that the shrinkage of a chain segment into a collapse blob does not
encounter a free-energy barrier. Accordingly, a collapse blob can not be
viewed as a critical droplet. Rather, the formation of a $\xi _{c}$-blob
resembles spinodal decomposition with the caveat that the associated
dissipation arises from configurational changes of the chain as a whole.
Thus, in order to minimize the entropy production the first step of the
collapse involves the formation of a small number $p$ ($\ll N/g_{c}$) of
collapse blobs instead of a string of $N/g_{c}$ $\xi _{c}$-blobs

The concave region in $F_{int}$ is also the origin of the force-law
governing the extension a collapsed globule. When the end-to-end distance of
the stretched globule is fixed, a dense globule coexists with a stretched
string of $\xi _{c}$-blobs. The tension $f_{co}$ associated with this
coexistence is the driving force for the third stage of the collapse. The
extension of the globule actually involves three 
regimes~\cite{HZ,foot}. 
In the first regime the spherical form of the globule is initially
deformed into a prolate ellipsoid, but maintains a constant volume
corresponding to close packing of the $\xi _{c}$-blobs. Within this
linear-response regime the free-energy penalty $F$ incurred is due to the
increase of the surface free energy 
\begin{equation}
F\approx \gamma \Delta A\approx \gamma (L-R_{c})^{2},  
\label{iii7}
\end{equation}
where $\Delta A\approx (L-R_{c})^{2}$ is the surface area increment
associated with the deformation, and $L$ is the length of the axis of
rotation. The corresponding restoring force, $f=-\partial F/\partial R$, 
is proportional to the strain $(L-R_{c}):$%
\begin{equation}
f\approx -\gamma (L-R_{c}).  \label{iii8}
\end{equation}
This type of process can not proceed indefinitely. If it were pursued
indefinitely, the distorted globule will assume a cylindrical shape and,
eventually, form a string of $\xi _{c}$-blobs. This scenario gives rise to a
van der Waals loop in the $f$ {\em vs.} $R$ diagram. In turn, this is
indicative of instability with respect to a coexistence of a weakly
elongated globule and a stretched string of $\xi _{c}$-blobs. This effect is
reminiscent of the Rayleigh-Plateau instability~\cite{R-P} involving the
break-up of a fluid jet into a succession of droplets. In the second,
coexistence, regime the chain comprises a stretched string of $n/g_{c}$ $\xi
_{c}$-blobs and a roughly spherical globule of $(N-n)/g_{c}$ closely packed $%
\xi _{c}$-blobs. To characterize this equilibrium situation it is necessary
to minimized the free energy of a stretched chain comprised of a dense
globule and a string of $n/g_{c}$ collapse blobs, i.e., 
\begin{equation}
\frac{F}{kT}\approx -\left( \frac{N-n}{g_{c}}\right) +\left( \frac{N-n}{g_{c}%
}\right) ^{2/3}+\frac{L^{2}}{na^{2}}.  \label{iii9}
\end{equation}
The first term allows for the transfer of the collapse blobs from the
solvent into the ``dense phase,\rlap\rq\rq\ i.e.,  the interior of the globule.
This is the primary driving term for the growth of the globule. The second
term reflects the surface free energy of the globule. The elastic free
energy of the string of blobs is described by the third term. Note that $%
na^{2}\approx (n/g_{c})\xi _{c}^{2}$ is the unperturbed span of an ideal
string of collapse blobs. Making $F$ stationary with respect to $n,$ $%
\partial F/\partial n=0,$ leads to 
\begin{equation}
\frac{1}{g_{c}}\left[ 1-\left( \frac{N-n}{g_{c}}\right) ^{-1/3}\right]
\approx \frac{1}{g_{c}}\approx \frac{L^{2}}{n^{2}a^{2}}.  \label{iii10}
\end{equation}
In the limit of large $L$ and for $n\lesssim N$ this leads to 
\begin{equation}
\frac{L}{\xi _{c}}\approx \frac{n}{g_{c}}.  \label{iii11}
\end{equation}
The equilibrium free energy of the resulting tadpole configuration is, up to
numerical prefactors

\begin{equation}
\frac{F}{kT}\approx \left( \frac{N-n}{g_{c}}\right) ^{2/3}+\left( \frac{n}{%
g_{c}}-\frac{N}{g_{c}}\right) +\frac{n}{g_{c}}.  \label{iii12}
\end{equation}
The first term allows for the surface free-energy $\gamma r_{g}^{2}$ of a
globule of radius $r_{globule}\approx \lbrack (N-n)/g_{c}]^{1/3}\xi _{c}$.
The transfer free-energy of blobs into the dense phase is reflected in the
second term. The third term is the stretching energy of the extended string
of length $L_{s}\approx (n/g_{c})\xi _{c}$ as given by the Gaussian
expression for the elastic penalty $kTL_{s}^{2}/na^{2}.$ It is also possible
to interpret this last term as the surface energy of the extended string.
The end-to-end distance reflects two contributions 
\begin{equation}
L\approx \xi _{c}\left[ \left( \frac{N-n}{g_{c}}\right) ^{1/3}+\frac{n}{g_{c}%
}\right] .  \label{iii13}
\end{equation}
The first term is the radius of the globule, and the second is the span of
the stretched string of $\xi _{c}$-blobs. When $n$ is sufficiently large we
may approximate $dR$ as $\xi _{c}dn/g_{c}$ and the corresponding force law $%
f=-\partial F/\partial L$ reads 
\begin{equation}
f/kT\approx -\xi _{c}^{-1}+r_{globule}^{-1}.  \label{iii14}
\end{equation}
It is important to note that $f$ decreases as $r_{globule}$ approaches $\xi
_{c}$.\ This finite-size correction is reminiscent of the Laplace law for
the vapor pressure of small droplets. It is negligible over a wide range of
extensions if $N$ is large enough. Because of this correction different
deformation scenarios are expected for the $f=const^{\prime }$ and the $%
L=const^{\prime }$ ensembles. In the first case the globule unravels
completely as soon as a critical force $f_{co}/kT\approx 1/\xi _{c}$ is
applied. No globule-coil coexistence is expected. Such a coexistence is
however expected when an end-to-end distance is imposed. In this case the
tension in the string is $f_{co}$. The onset of the coexistence occurs when $%
L\approx R_{c}+\xi _{c}$ as can be seen by equating $f_{co}/kT$ to the force
given by Eq. (\ref{iii8}). The upper boundary of this second regime
corresponds to a fully extended string of blobs, $L_{\max }\approx
(N/g_{c})\xi _{c}.$ Stronger extension is characterized by simple Gaussian
elasticity, up to the onset of finite extensibility effects.

\section{Collapse to $\protect\theta $ Conditions and the Formation of
Nascent Droplets}

\begin{figure}[hbt]
 \epsfxsize=3.5in
 \centerline{\epsfbox{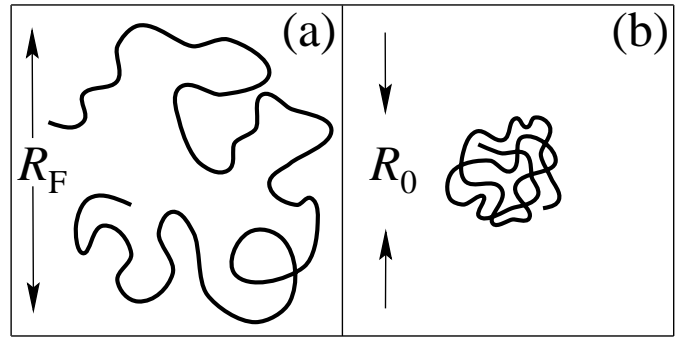}}
\vskip0.5truecm
\caption{Collapse to $\theta$ conditions: 
\qquad\qquad\qquad\qquad\qquad\qquad
\\
(a)~initial, self-avoiding configuration; 
\\
(b)~the final, random walk configuration.}
\label{FIG:figtwo}
\end{figure}

A quench from $T>\theta $ to $T=\theta $ causes the chain to shrink from $%
R_{F}/a\approx v^{1/5}N^{3/5}$ to $R_{o}/a\approx N^{1/2}$ 
(Fig.~\ref{FIG:figtwo}).
The dynamics of this process are of interest for two reasons. First, as we shall
discuss, the formation of nascent pearls follows similar kinetics. Second,
the discussion of this process is especially simple, thus providing a useful
introduction to the subsequent sections. The quench to $T=\theta $ turns off
the repulsive interactions between the monomers so that driving force for
the ``collapse to $\theta "$ is due\ solely to the Gaussian elastic free
energy of the chain, $F_{chain}\approx F_{el}$ or 
\begin{equation}
\frac{F_{el}}{kT}\approx \alpha ^{2}-\ln \alpha ,  \label{iv1}
\end{equation}
where $\alpha =R/R_{o}$ and $R=R(t)$ is the instantaneous chain radius at
time $t$. To obtain the characteristic time for this process we equate the
time derivative of the elastic energy, 
\begin{equation}
\frac{dF_{el}}{dt}\approx \left( \alpha -\frac{1}{\alpha }\right) \frac{%
d\alpha }{dt},  \label{iv2}
\end{equation}
to the dissipative losses associated with the shrinking, $-TdS_{total}/dt,$
where $dS_{total}/dt$ is the corresponding entropy production. Two
dissipation processes are involved. One is the viscous dissipation due to
the hydrodynamic flow resulting from the contraction of the coil. As was
discussed previously, the resulting entropy production within the coil is
roughly 
\begin{equation}
T\frac{dS_{h}}{dt}\approx \eta V^{2}R,  \label{iv3}
\end{equation}
where $V=dR/dt$ is the velocity of the flow. Second is the entropy
production due to the Stokes drag force experienced by the monomers as they
move through the solvent. The associated entropy production is approximately 
\begin{equation}
T\frac{dS_{S}}{dt}\approx \eta NaV^{2}.  \label{iv4}
\end{equation}
\noindent
The dynamics of the collapse are determined by $%
dF_{el}/dt=-T(dS_{h}/dt+dS_{S}/dt).$ Whilst in this case it is possible to
solve the full equation, it is easier, and more instructive, to consider $%
dF_{el}/dt=-TdS_{h}/dt$ and $dF_{el}/dt=-TdS_{S}/dt$ separately. $%
dF_{el}/dt=-TdS_{h}/dt$ leads to 
\begin{equation}
\frac{dt}{\tau _{Z}}\approx \frac{\alpha ^{2}d\alpha }{\alpha ^{2}-1},
\label{iv5}
\end{equation}
and the characteristic time associated with the hydrodynamic dissipation is
thus the Zimm time~\cite{PGG} of the ideal coil, i.e., 
\begin{equation}
\tau _{Z}\approx \frac{\eta R_{o}^{3}}{kT}\sim N^{3/2}.  \label{iv6}
\end{equation}
The frictional dissipation is dominant as it scales with $N$ rather than $R.$
It leads to 
\begin{equation}
\frac{dt}{N^{1/2}\tau _{Z}}\approx \frac{\alpha d\alpha }{\alpha ^{2}-1},
\label{iv7}
\end{equation}
and the corresponding characteristic time is the Rouse relaxation time\cite
{PGG} of an ideal coil $\tau _{R}\approx \eta N^{2}a^{3}/kT$ or 
\begin{equation}
\tau _{R}\approx \tau _{Z}N^{1/2}\sim N^{2}.  \label{iv8}
\end{equation}
As the Rouse time is much longer, it dominates the process.

Within our picture, the first stage of the collapse is the formation of
nascent pearls comprising each of roughly $g_{c}$ monomers. The process
involves$\ p$ randomly distributed chain segments. Initially, each of the
chain segments exhibits self-avoidance. Once the solvent is quenched, the
monomer-monomer repulsions disappear and the segments shrink to their ideal
coil configuration. As the nascent pearls consist of only $g_{c}$ monomers
the monomer-monomer attraction is not strong enough to perturb the ideal
coil configuration within them. So long as $p$ is small enough, the overall
radius of the chain is only weakly modified and the dissipation is due to
the local configurational changes. In this picture it is possible to neglect
the couplings between the droplets and between the droplets and the chain as
whole. Thus, the dynamics of this process are essentially identical to those
of the collapse of a chain comprising $g_{c}$ monomers to $\theta $
conditions. The characteristic time is, accordingly the corresponding Rouse
time or 
\begin{equation}
\tau _{p}\approx \tau _{\xi }g_{c}^{1/2}\approx \frac{\eta a^{3}}{kT}%
g_{c}^{2}\sim \left( \frac{\theta }{\Delta T}\right) ^{4},  \label{iv9}
\end{equation}
where 
\begin{equation}
\tau _{\xi }\approx \frac{\eta \xi _{c}^{3}}{kT}\sim \left( \frac{\theta }{%
\Delta T}\right) ^{3},  \label{iv10}
\end{equation}
is the Zimm time of a collapse blob. Note that $\tau _{p}$ is independent of 
$N$ and $p$ as expected for a local process. A short relaxation time \
scaling as $N^{o}$ is also predicted by Langevin type theories of 
collapse~\cite{Dawson,Garel}.

\section{Collapse of a Chain with Constrained Ends: The
Bridge-Stretching Regime}

\begin{figure}[hbt]
 \epsfxsize=3.5in
 \centerline{\epsfbox{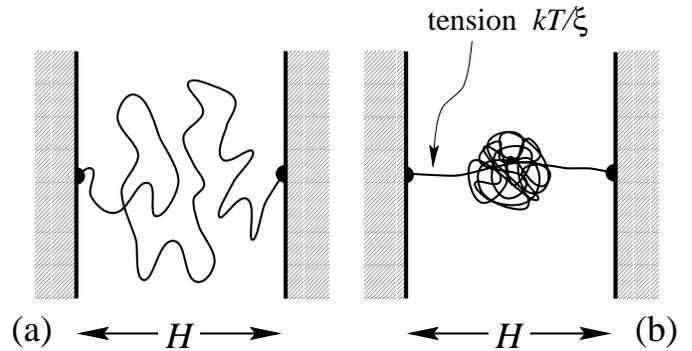}}
\vskip0.5truecm
\caption{Collapse of chain with constrained ends: 
\\
(a)~the initial, swollen configuration; 
\\
(b)~the final ``tadpole'' configuration in a poor solvent.}
\label{FIG:figthr}
\end{figure}

In the second stage of the collapse, the roughly stationary droplets,
``pearls'', grow by ``sucking in'' monomers from the bridges joining them.
This stage ends when the initially undeformed bridges assume the
configuration of a stretched string of $\xi _{c}$-blobs. In comparison to
the ``spinodal'' stage described in Section IV, this process is
distinguished by two features. First, the driving force is now due to the
transfer free energy of the free monomers into the dense phase. Second, the
evolution of the droplets at this stage strongly affects the configuration
of the bridges. The bridges stretch as the number of monomers in them
decreases. We first consider the somewhat simpler situation of the collapse
of a chain with its ends constrained to fixed positions separated by a
distance $L\gg R_{c}$ 
(Fig.~\ref{FIG:figthr}). 
The free energy of such a chain, neglecting
the surface free-energy, is given by 
\begin{equation}
\frac{F_{chain}}{kT}\approx -\left( \frac{N-n}{g_{c}}\right) +\frac{L^{2}}{%
na^{2}},  \label{v1}
\end{equation}
where $n$ is the number of ``free'' monomers that have not been assimilated
into a globule. We assume that the $n$ monomers behave as a random walk, 
i.e., the instantaneous span of the chain perpendicular to the end-to-end
vector is 
\begin{equation}
R_{\bot }\approx (n/g_{c})^{1/2}\xi _{c}\approx n^{1/2}a,  \label{v2}
\end{equation}
and as a result 
\begin{equation}
\frac{dn}{dt}\approx \frac{1}{a^{2}}R_{\bot }\frac{dR_{\bot }}{dt}.
\label{v3}
\end{equation}
\noindent
Relations~(\ref{v2}) and (\ref{v3}) are actually invalid at the initial
stages of this regime, when the configurations of the chain still exhibit
self-avoidance. They are, however, expected to apply at the final stages,
which dominate the relaxation time. Invoking them implies that the
relaxation of the chain configuration when $n$ decreases is effectively
instantaneous. The driving force for this stage, as specified by Eqs. (\ref
{v1})-(\ref{v3}), is 
\begin{equation}
\frac{1}{kT}\frac{dF_{chain}}{dt}\approx \left( \frac{1}{g_{c}}-\frac{L^{2}}{%
n^{2}a^{2}}\right) \frac{1}{a^{2}}R_{\bot }\frac{dR_{\bot }}{dt}.  \label{v4}
\end{equation}
The driving force is opposed by the two dissipation mechanism invoked in
Section IV. In the present situation, however, the dissipation reflects the
cylindrical symmetry of the system. The hydrodynamic dissipation, due to the
flow of the solvent as it is expelled from the region occupied by the coil,
is $T(dS_{h}/dt)\approx \eta (dR_{\bot }/dt)^{2}L$. The Stokes drag on the
bridges formed by the collapsed blobs is $T(dS_{b}/dt)\approx \eta
(n/g_{c})\xi _{c}(dR_{\bot }/dt)^{2}.$ The dynamics of the collapse are
determined by $dF_{chain}/dt=-T(dS_{total}/dt).$ At equilibrium, $n$ is $%
n_{eq}\approx (L/\xi _{c})g_{c}\approx g_{c}^{1/2}(L/a).$ Introducing the
variable $y=n/n_{eq}$ we obtain 
\begin{equation}
\frac{\xi _{c}}{L}\frac{dt}{\tau _{\xi }}\approx -\frac{ydy}{y-1},
\label{v5}
\end{equation}
and the characteristic time for this collapse process is thus 
\begin{equation}
\tau \approx \frac{L}{\xi _{c}}\tau _{\xi }.  \label{v6}
\end{equation}

A similar scenario occurs in the bridge-stretching regime 
(Fig.~\ref{FIG:figone}). 
In this stage the $p$ randomly placed pearls are roughly stationary, and they
grow by assimilating monomers from the bridges joining them. The bridges are
initially undeformed. This stage ends when the bridges consist of a
stretched string of collapse blobs. The average length of each stretched
bridge is $R_{p}\approx (n/g_{c}p)\xi _{c}.$ As, within this model, the $p$
bridges are randomly placed, the overall span of the chain is $R\approx
p^{1/2}R_{p}.$ By assuming that at this stage the overall size of the chain
remains $R\approx R_{F}\approx N^{3/5}a$ and that $n\approx N$ we are led to
the result 
\begin{equation}
p\approx N^{4/5}/g_{c}.  \label{v6a}
\end{equation}
This amounts to the completion of all short length-scale process prior to
the onset of the overall shrinking of the chain. The growth of the
dissipation with the length-scale involved justifies this estimate. Clearly,
this is a rough estimate and, in reality, $R$ is expected to shrink
somewhat. The estimated value of $p,$ Eq. (\ref{v6a}), specifies $R_{p}$ at
the end of the stretching stage to be 
\begin{equation}
R_{p}\approx N^{1/5}\xi _{c}  \label{v7}
\end{equation}
and the corresponding characteristic time is than 
\begin{equation}
\tau _{bs}\approx N^{1/5}\tau _{\xi }  \label{v8}
\end{equation}
In contrast with $\tau _{p}$, the time scale $\tau _{bs},$ depends on both $p
$ and $N.$

\section{Collapse of the ``Pearl Necklace''}

At the end of the bridge-stretching regime, the configuration of the chain
is that of a random walk of $p$ steps of length $R_{p}$ and with an overall
radius $R_{F}\sim N^{3/5}$. The $p\approx N^{4/5}/g_{c\text{ }}$elementary
steps are dumb-bells comprising of two pearls joined by a stretched string
of\ $\xi _{c}$-blobs. It is impossible to incorporate more monomers into the
pearls while keeping them stationary. The pearls can only grow by shortening
the bridges, thus causing the overall shrinking of the chain from $R\sim
R_{F}$ to $R\approx R_{c}\sim N^{1/3}.$ It is assumed that the pearls are
the vertices of a random walk and that their number, $p$, is constant. This
approximation is reasonable in the early phase of this stage but may be
questionable at later times. This issue will be discussed later. On short
time-scales the bridges equilibrate rapidly whereas the pearls remain
essentially fixed in space. Consequently, the positioning of the pearls does
not give rise to an entropy and the free energy corresponding to this
process is $F\approx pF_{pb},$ where\ $F_{pb}$ is the free energy of a pearl
attached to a stretched string of collapse blobs. In turn, $F_{pb}$
comprises two terms. One is the excess free energy of a ``free'' collapse
blob as given by the $kT$ per blob Ansatz. The second is the elastic free
energy of the stretched bridge of length $R_{p}$ with respect to its
reference state, i.e., a random walk of collapse blobs. The span of an
undeformed bridge with an ideal coil configuration is $R_{p0}\approx
(n/pg_{c})^{1/2}\xi _{c}$ and the associated Gaussian stretching penalty is $%
kT(R_{p}/R_{p0})^{2}.$ Assuming that the overall configuration is that of a
random walk, $R\approx p^{1/2}R_{p},$ the free energy of the chain as a
whole is given by 
\begin{equation}
\frac{F_{chain}}{kT}\approx p\left( -\frac{N-n}{pg_{c}}+\frac{R^{2}}{na^{2}}%
\right) .  \label{vi1}
\end{equation}
The equilibrium state of the chain, at a given instant, is specified by $%
\partial F/\partial n=0$ for constant $R$ and $p,$ thus leading to 
\begin{equation}
R\approx \frac{na}{\sqrt{pg_{c}}}.  \label{vi2}
\end{equation}
We assumed that the decrease in $R$ is slow in comparison to the relaxation
of $n$ for a given $R.$ Accordingly, the free energy for a certain $R,$ as
obtained from Eqs. (\ref{vi1}) and (\ref{vi2}) is given by 
\begin{equation}
\frac{F_{chain}}{kT}\approx -\frac{N}{g_{c}}+\sqrt{p}\frac{R}{\xi _{c}},
\label{vi3}
\end{equation}
and the corresponding force, $f/kT\approx \partial F/\partial R,$ is 
\begin{equation}
f\approx \sqrt{p}\frac{kT}{\xi _{c}}\approx \sqrt{p}f_{co},  \label{vi4}
\end{equation}
where $f_{co}\approx kT/\xi _{c}$ is the equilibrium tension in the
dumb-bell shaped, coexistence configuration. If the string of pearls is
perfectly collinear, $R\approx pR_{p},$ similar considerations lead to $%
R\approx (n/g_{c})\xi _{c}$ and $f\approx f_{co}$. In this situation only
the terminal pearls experience a net force. As the tension in the bridges,
ignoring finite size effects, is $f_{co},$ the net force on the middle
pearls is zero. However, if the pearls are the vertices of a random walk,
each pearl is subject to a random force $\sim f_{co}$ thus leading to (\ref
{vi4}). The thermodynamic driving force for the collapse is simply $%
dF_{chain}/dt\approx fdR/dt.$ This reflects the rate of change in the number
of ``free'' $\xi _{c}$-blobs, $d(n/g_{c})/dt.$ As before, two dissipation
mechanisms are involved. The hydrodynamic dissipation is $%
T(dS_{h}/dt)\approx \eta (dR/dt)^{2}R.$ The dissipation modes in pearl
necklace situation differs from the stretched-bridges case in two respects.
First, $R(t)$ replaces $L\approx const^{\prime }$ because the coil is
assumed to shrink as a sphere. Second, as the pearls are no longer
stationary it is necessary to consider two dissipation modes associated with
the Stokes drag force. One is due, as before, to the contribution of the
stretched bridges whereas the second arises from the motion of the
hydrodynamically impenetrable pearls. The pearls term is $%
T(dS_{p}/dt)\approx p\eta r_{p}(dR/dt)^{2},$ where the volume of a pearl of
closed packed $\xi _{c}$-blobs, $r_{p}^{3}\approx (N-n)/pg_{c},$ specifies
the radius of an individual pearl. The bridges term is, similarly, $%
T(dS_{b}/dt)\approx p\eta R_{p}(dR/dt)^{2},$ where $R_{p}\approx
(n/g_{c})\xi _{c\text{ }}$ is the length of an individual bridge.
Altogether, $dF_{chain}/dt=-T(dS_{total}/dt)$ leads to 
\begin{equation}
pg_{c}\frac{dt}{\tau _{\xi }}\approx -\left[ \frac{n}{\sqrt{p}g_{c}}+p\left( 
\frac{N-n}{pg_{c}}\right) ^{1/3}+\frac{n}{g_{c}}\right] dn,  \label{vi5}
\end{equation}
where the three terms on the right hand side correspond, respectively, to
the dissipation due to hydrodynamic flow, the pearls and the bridges. In
this case it is simpler to solve individually each $%
dF_{chain}/dt=-T(dS_{i}/dt),$ in order to find the characteristic time for
the different dissipation modes and then to identify the dominant one. The
hydrodynamic dissipation yields 
\begin{equation}
\tau _{h}\approx \left( \frac{N}{g_{c}}\right) ^{2}p^{-3/2}\tau _{\xi },
\label{vi6}
\end{equation}
whereas the bridges contribution leads to 
\begin{equation}
\tau _{b}\approx \left( \frac{N}{g_{c}}\right) p^{-1}\tau _{\xi },
\label{vi7}
\end{equation}
and the pearls dissipation to 
\begin{equation}
\tau _{pd}\approx \left( \frac{N}{g_{c}}\right) ^{4/3}p^{-1/3}\tau _{\xi }.
\label{vi8}
\end{equation}
Estimating $p$ by $p\approx N^{4/5}/g_{c}$ yields $\tau _{h}\sim
N^{4/5}/g_{c}^{1/2},$ $\tau _{b}\sim N^{6/5}/g_{c},$ and $\tau _{pd}\sim
N^{16/15}/g_{c}$ leading to the identification of $\tau _{b}\approx
(N^{6/5}/g_{c})\tau _{\xi }$ as the longest relaxation time and the bridges
dissipation as the rate-determining dissipation mode. Note, however, that
this is the case only for sufficiently deep quenches, when $g_{c}<N^{2/5}.$
By contrast, hydrodynamic dissipation is dominant when $g_{c}>N^{2/5}.$

The discussion presented above is based on three assumptions: (i)
Finite-size corrections are negligible and $f_{co}\approx kT/\xi _{c}.$ (ii)
The pearls are placed at the vertices of a random walk (iii) $p\gg 1$ and $%
p\approx const^{\prime }$. The first assumption is easiest to justify. The
finite size correction to $f_{co}$ is $1/r_{p},$ where $r_{p}$ is the radius
of the pearl. This correction is expected to be small at the end of the
bridge-stretching stage. In any case, it decreases with time. The second
assumption is reasonable, within our model, at the onset of the collapse of
the pearl necklace. Its range of validity is however uncertain. Some
simulation results support this picture~\cite{Byme} whilst others~\cite{Dawson1}
suggest that the chain may approach a linear configuration. The assumption
that $p\gg 1$ is reasonable for $N\gg 1.$ Both~\cite{Byme} and~\cite{Dawson1}
support this view. The assumption of $p\approx const^{\prime }$\ is clearly
a rough approximation and $p$ is expected to decrease with time. It is
nevertheless interesting to note that both~\cite{Byme} and~\cite{Dawson1}
suggest that $p$ varies very weakly over a wide range of times.
``Pearl-pearl'' coalescence is the most likely mechanism for the decrease in 
$p.$ Initially, this is due to one-dimensional diffusion of the pearls that
coalesce upon contact because of the surface free energy. At this point it
is useful to consider two possible scenarios for the coalescence of the
pearls. In the first, pearl diffusion is fast in comparison to the overall
collapse of the chain. In this case the one dimensional diffusion incurs no
free-energy penalties as long as $n$ is constant. The favored mode thus
involves the transfer of monomers between two adjacent bridges and through
the adjoining pearl. This transfer of monomers across the pearl must take
place along the trajectory of the chain. It is, thus, a reptation type
motion in the sense that the monomer needs to diffuse across a distance $%
(r_{p}/a)^{3}a$ instead of $r_{p}.$ This dissipation mode may well dominate
over the Stokes drag force. Only the terminal pearls may diffuse in a
non-reptative mode. In the other limit the pearl diffusion is relatively
fast. In this case it can result from asymmetry in the rate of assimilating
monomers from the two bridges. In this situation, the diffusion coefficient
will reflect the Stokes drag force on the pearls. In any case, the dynamics
of diffusion-controlled coalescence in one-dimensional is complex. The case
of the $A+A\rightarrow A$ reaction, with constant diffusion coefficient $D,$
was studied extensively~\cite{PDF}, and a rigorous solution was reported
only recently~\cite{bA}. At short times the linear concentration, $c,$
follows the mean-field rate equation $dc/dt\varpropto -c^{2}$. However, the
asymptotic solution at large times is $c\sim 1/\sqrt{Dt}$ instead of $c\sim
1/Dt,$ as predicted by the mean-field treatment. The situation encountered
in the present case is more complicated in two respects: First, the
diffusion coefficient is no longer a constant. Rather, it depends on $r_{p}$
and thus on $t.$\ Second, the total length varies with $p$ and $r_{p}$ thus
introducing a novel mechanism for the $t$ dependence of $c.$ A complete
analysis of this problem is beyond the scope of the present paper. It is
however possible to estimate a characteristic time, $\tau _{pp},$ for the
onset of coalescence. $\tau _{pp}$ may be identified with the time required
by a pearl in order to diffuse across a distance $R_{p}\approx N^{1/5}\xi
_{c},$ as specified by the Einstein diffusion relation $D_{p}\tau
_{pp}\approx R_{p}^{2}.$ A lower bound for $\tau _{pp}$ is obtained by
assuming that the diffusion coefficient of the pearl scales as $D_{p}\sim
1/\xi _{c}$ thus leading to $\tau _{pp}\sim N^{2/5}.$

\section{Concluding Remarks}

The literature concerning the theory of the dynamics of collapse describes
three different approaches: Langevin-type theories, simulation studies and
phenomenological models. It is simplest to compare the present model to the
earlier phenomenological models. The kinetics of collapse were first
considered by de Gennes in 1985~\cite{PGG1}. In this model the collapsing
chain was viewed as a ``sausage'' of collapse blobs. With time, the sausage
diameter increases and its length decreases. The driving free energy was
identified with the interfacial free energy and the dissipation was
attributed to hydrodynamic flow. A different view was later proposed by
Buguin, Brochard-Wyart and de Gennes~\cite{Beguin}. Within this model the
driving free energy is the transfer free energy into the dense phase. Two
scenarios were considered for the early collapse. In one, the pearls form
the vertices of a random walk. This scenario differs from the present model
in two respects. First, the bridges joining the pearls are assumed to be
unstretched. Second, the associated dissipation arises only from the
hydrodynamic flow. Importantly, the collapse time within this picture does
not depend on the number of pearls, $p.$ In the second scenario, the chain
is envisioned as a dumb-bell consisting of two pearls, $p=2,$ joined by a
stretched string of collapse blobs. The dissipation in this case is
attributed to the Stokes drag force on the pearls. Most recently the problem
was re-examined by Klushin~\cite{Klushin}. In this model the driving free
energy is the transfer free energy to the dense phase and the dissipation is
due to the Stokes drag force. The chain is envisioned as a self-avoiding
random walk of Kuhn lengths. In turn, these can comprise either stretched
bridges joining pearls or by a sequence of collinear pearls connected by
stretched bridges. Thus, the number of Kuhn lengths can change either
because of coalescence or because of the formation of a collinear sequence.
The rate of change in the number of Kuhn lengths is assumed to follow
unimolecular kinetics. The characteristic time for this process is
identified with that of a shrinking dumb-bell, assuming a constant size of
the pearls. The number of pearls within a collinear sequence is assumed to
decrease because monomer transfer or coalescence, both driven by the surface
free-energy. Within this model, the collapse is largely due to the decrease
in $p$. The models of Buguin et al.~and of Klushin focus, in effect,
on the kinetics of the collapse of a pearl necklace and do not consider
possible earlier events.

In principle, the Langevin equation can provide a complete description of
the collapse on all time-scales and thus allows for a systematic analysis of
the problem. Unfortunately, a rigorous and complete solution of the
appropriate Langevin equation is difficult. Accordingly, all the
Langevin-type theories introduce certain approximations so as to reach a
mathematically tractable formalism. The treatment is mathematically
demanding, and often requires numerical calculations. A detailed comparison
with phenomenological theories is difficult because of the very different
formulations. It is interesting however to note that both the theory of
Garel {\em at al.}~\cite{Garel} and of Kuzentsov et al.~\cite{Dawson}
distinguish between two regimes in the early collapse: first, a very fast
process, whose characteristic time is independent of $N$ and is thus
interpreted as local rearrangement; and second, a much slower process,
reflecting large scale reorganization, with a characteristic time that
increases with $N.$

We hope that a full picture of the kinetics of collapse will eventually
emerge from molecular dynamics simulations involving a polymer chain as well
as solvent molecules. The currently available simulation studies do not
account fully for hydrodynamics thus introducing an uncertainty in the
interpretation of the results. In certain studies the interpretation is
further hampered by the imposition of an ``all trans'' 
initial state~\cite{KS,Mattice}. 
With this caveat, the simulations mostly support the notion
that the collapse involves a string of pearls joined by stretched bridges as
an intermediate stage. The configuration adopted by the string of pearls is
less clear. Snapshots of the chain configuration as obtained by
Langevin-equation simulation of Byme et al.~\cite{Byme} suggest an
almost collinear string of pearls. The visualization of configurations
obtained by Monte Carlo simulation of Kuznetsov et al.~\cite{Dawson1}
reveal a random walk pearl necklace. These last two studies also suggest
that the first step of the collapse involves the formation of localized
clusters, and that subsequent cluster growth is attained by assimilating
monomers from the bridges thus causing stretching.

\noindent
{\it Acknowledgments\/}: 
The authors have benefited from extensive discussions with A.~Yu.~Grosberg.  Support from NSF-INT96-03228 (A.H., P.M.G.), 
CNRS-NSF US-France Cooperative Research Grant (A.H.), and 
NSF-DMR99-75187 (P.M.G.) is gratefully acknowledged.

%
%
%
\end{multicols}
\end{document}